\documentclass{PoS}

\title{LOFAR Transients and the Radio Sky Monitor}

\ShortTitle{LOFAR Transients}

\author{\speaker{Rob Fender},$^{ab}$ Ralph Wijers,$^b$ Ben
         Stappers,$^{bc}$ Robert Braun,$^c$ Michael Wise,$^{bc}$ Thijs
         Coenen,$^{b}$ Heino Falcke,$^{cd}$ Jean-Mathias
         Griessmeier,$^{e,c}$ Michiel van Haarlem,$^c$ Ger de Bruyn,$^c$,
         Peter Jonker,$^{fgh}$ Casey Law,$^b$ Sera Markoff,$^b$ Joseph
         Masters,$^b$ James Miller-Jones,$^b$ Rachel Osten,$^i$ Bart
         Scheers,$^b$ Hanno Spreeuw,$^b$ John Swinbank,$^b$ Corina
         Vogt,$^c$ Rudy Wijnands$^b$ and Philippe Zarka,$^e$\\
         \llap{$^a$}School of Physics and Astronomy, University of
         Southampton, Highfield, Southampton, SO17 1BJ, UK\\
         \llap{$^b$}Astronomical Institute `Anton Pannekoek',
         University of Amsterdam, Kruislaan 403, 1098 SJ, Amsterdam,
         the Netherlands\\ \llap{$^c$}Stichting ASTRON, Postbus 2,
         7990 AA Dwingeloo, the Netherlands\\ \llap{$^d$}Department of
         Astronomy, Radboud University, Postbus 9010, 6500 GL
         Nijmegen, the Netherlands\\ \llap{$^e$}Observatoire de
         Paris-Meudon, 5 Place Jules Janssen, 92195 Meudon Cedex,
         France\\ \llap{$^f$}SRON, Netherlands Institute for Space
         Research, Sorbonnelaan 2, 3584 CA, Utrecht, the Netherlands\\
         \llap{$^g$}Harvard-Smithsonian Center for Astrophysics, 60
         Garden Street, Cambridge, MA 02138, USA\\
         \llap{$^h$}Astronomical Institute, Utrect University, Postbus
         80000, 3508 TA, Utrecht, the Netherlands\\
         \llap{$^i$}Department of Astronomy, University of Maryland,
         College Park, MD, USA\\ E-mail: \email{rpf@phys.soton.ac.uk},
         \email{rwijers@science.uva.nl},\email{ben.stappers@manchester.ac.uk},
         \email{robert.braun@csiro.au}, \email{wise@science.uva.nl},
         \email{tcoenen@science.uva.nl}, \email{falcke@astron.nl},
         \email{griessmeier@astron.nl},
         \email{haarlem@astron.nl}, \email{p.jonker@sron.nl},
         \email{claw@science.uva.nl}, \email{sera@science.uva.nl},
         \email{jmasters@science.uva.nl},
         \email{jmiller@science.uva.nl}, \email{rosten@astro.umd.edu},
         \email{bscheers@science.uva.nl},
         \email{hspreeuw@science.uva.nl},
         \email{swinbank@science.uva.nl} \email{vogt@astron.nl},
         \email{rudy@science.uva.nl}, \email{philippe.zarka@obspm.fr}}

\abstract{ The study of transient and variable low-frequency radio
  sources is a key goal for LOFAR, with an extremely broad science
  case ranging from relativistic jets sources to pulsars, exoplanets,
  radio bursts at cosmological distances, the identification of
  gravitational wave sources and even SETI. In this paper we will very
  briefly summarize the science of the LOFAR Transients key science
  project, will outline the capabilities of LOFAR for transient
  studies, and introduce the LOFAR Radio Sky Monitor, a proposed mode
  in which LOFAR regularly scans $\sim 2\pi$ radians of sky.}
\FullConference{...}
\begin{document}

\section{Introduction}

LOFAR is a next-generation radio telescope under construction in The
Netherlands with long-baseline stations under development in other
European countries (currently Germany, The UK, France, Sweden). The
array will operate in the 30--80 and 120--240 MHz bands (80--120 MHz
being dominated by FM radio transmissions in northern Europe). The
telescope is the flagship project for ASTRON, and is the largest of
the pathfinders for the lowest-frequency component of the Square
Kilometere Array (SKA). Core Station One (CS1; see Gunst et al. 2006)
is currently operating, and the next stage of deployment is about to
begin, with 36 stations to be in the field by the end of 2009.

For more information on the project, see

\smallskip
{\bf http://www.lofar.org}
\smallskip

The science of LOFAR has been driven intially by four, and more
recently six, Key Science Projects (KSPs). These are:

\begin{itemize}
\item{{\bf The Epoch of Reionization:} LOFAR will detect the signal of
the 21 cm HI line redshifted to the era, a billion years after the big
bang, when the first sources of radiation (Pop III stars and the first
accreting black holes) began to reionize the Universe.}
\item{{\bf Extragalactic Surveys:} LOFAR will conduct the most
extensive surveys of extragalactic radio sources (AGN and starburst
galaxies) ever undertaken, not to be surpassed until the completion of
the SKA.}
\item{{\bf Transients and pulsars:} LOFAR will revolutionize the study
of bursting and transient radio phenomena, including pulsars, due to
its enormous field of view and the very low frequencies which maximise
sensitivity to coherent radio bursts.}
\item{{\bf Cosmic rays:} LOFAR will detect {\em geosynchrotron}
radiation from cosmic ray air showers, and will attempt to detect the
highest particles energies ever measured by looking for their
interaction with the lunar regolith.}
\item{{\bf Solar physics:} LOFAR will study the explosive particle
acceleration in the Sun's atmosphere, and the propagation of ejecta
through the inner solar system.}
\item{{\bf Cosmic Magnetism:} LOFAR will measure the ordering of
magnetic fields on all scales from our own galactic plane to galaxy
clusters, via deep measurements of polarisation.}
\end{itemize}

In addition to these KSPs, there will be an increasing fraction of
open time available to the wider user community. In this brief paper,
we highlight the science and implementation of the Transients KSP, and
in particular the very wide-field scan mode which we call the {\em
Radio~Sky~Monitor} (RSM).

\section{The LOFAR Transients Key Science Project}

The LOFAR Transients Key Science Project (KSP) aims to study all
variable and transients sources detected by LOFAR. Such sources can be
broadly divided into two types:

\begin{enumerate}
\item {\bf Incoherent synchrotron emission:} essentially all explosive
  events which inject energy into the ambient medium result in
  particle acceleration and/or compression/enhancement of magnetic
  fields, resulting in synchrotron emission. Such emission is likely
  to be initially self-absorbed at LOFAR frequencies, with a rise time
  corresponding to the timescale for the source to expand and become
  optically thin in a given band. This expansion timescale is
  proportional to the initial size of the emitting region divided by
  the expansion velocity. In approximate order of increasing rise
  time, sources which will be associated with such emission are jets
  from CVs, X-ray binaries, GRBs and SNe, and finally AGN. Under most
  conditions, synchrotron emission has a maximum brightness
  temperature of $\sim 10^{12}$K.  An example of an outburst from a
  galactic binary system is given in Fig 1 below.

\begin{figure}[h]
\includegraphics[width=.99\textwidth, angle=0]{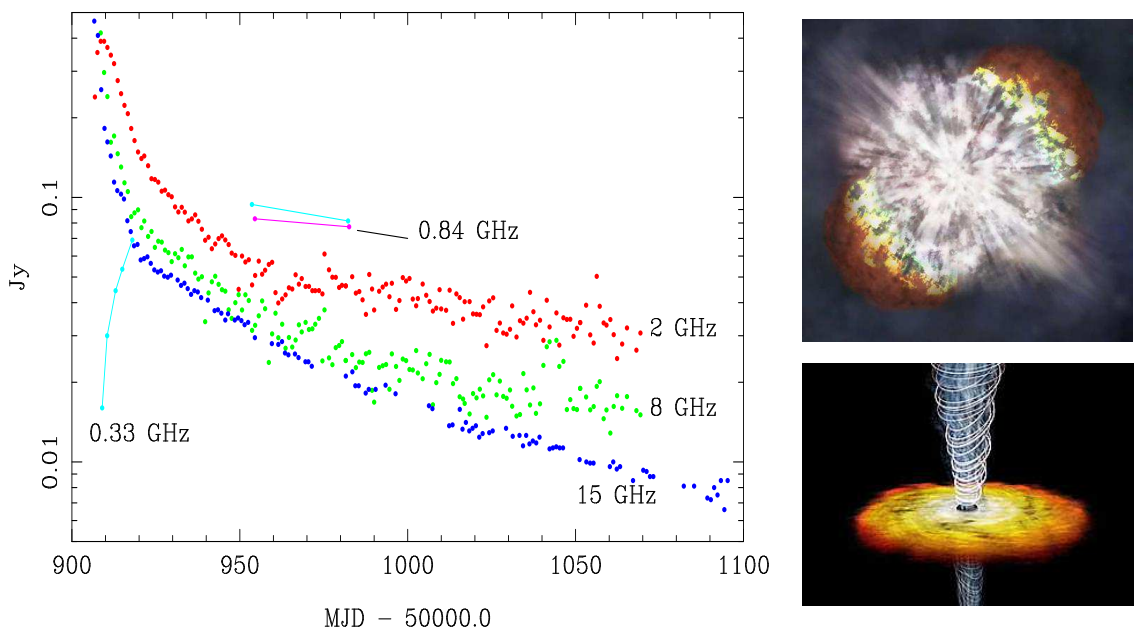}
\caption{Evolution of a radio outburst from the X-ray transient CI
  Cam, with data from the Ryle Telescope (15 GHz), the Green Bank
  Interferometer (2 and 8 GHz) and the Westerbork Synthesis Radio
  Telescope (0.8 and 0.3 GHz).  The radio signal is due to synchrotron
  emission from an expanding source of accelerated electrons. By the
  time of the first radio observations (`externally triggered' by
  X-ray observations) the radio source is already optically thin at
  the highest frequency, 15 GHz. As the source expands, the flux
  density at different frequencies increases as the optical depth at
  that frequency decreases, but then subsequently decays once in the
  optically thin regime as expansion results in energy losses. The
  emission at 330 MHz, just above the LOFAR band, was detectable
  within a few days of the outburst, but did not peak until 20--30
  days later. Such behaviour will be characteristic of explosive
  outburst events associated with e.g. relativistic jets, supernovae,
  GRB afterglows (as indicated by the cartoons to the right of the
  figure), with the rise and decay timescales being an increasing
  function of the luminosity of the event.}
\end{figure}

Note that many of the types of object responsible for explosive
particle acceleration, e.g. X-ray binaries, Gamma-ray bursts, are of
intense interest to the high-energy astrophysics. LOFAR detections of
such events, especially with the Radio Sky Monitor (see below), will
undoubtedly therefore be used as triggers for optical / infrared /
X-ray / gamma-ray follow-up (note for a typical X-ray binary outburst,
the majority of the interesting X-ray behavior would still be observed
even if the LOFAR trigger was $\sim 20$ days after the initial
outburst due to the optical depth delay described above). Taken as an
ensemble, observations of the synchrotron sources will provide a
complete time-resolved census of particle acceleration in the local
universe, shedding light on the energization of ambient media and
sites of cosmic ray acceleration.

\item {\bf Coherent emission:} several different types of radio
  emission with brightness temperatures $\geq 10^{12}$K, often
  resulting from very short durations, are lumped together under the
  title of `coherent' radio emission. This is generally taken to mean
  groups of electrons moving together {\em en masse}, and may often be
  highly anisotropic (e.g. maser emission). Several different classes
  of coherent radio emitters are likely to be detected by LOFAR, e.g.
\begin{itemize}

\begin{figure}[h]
\includegraphics[width=.99\textwidth]{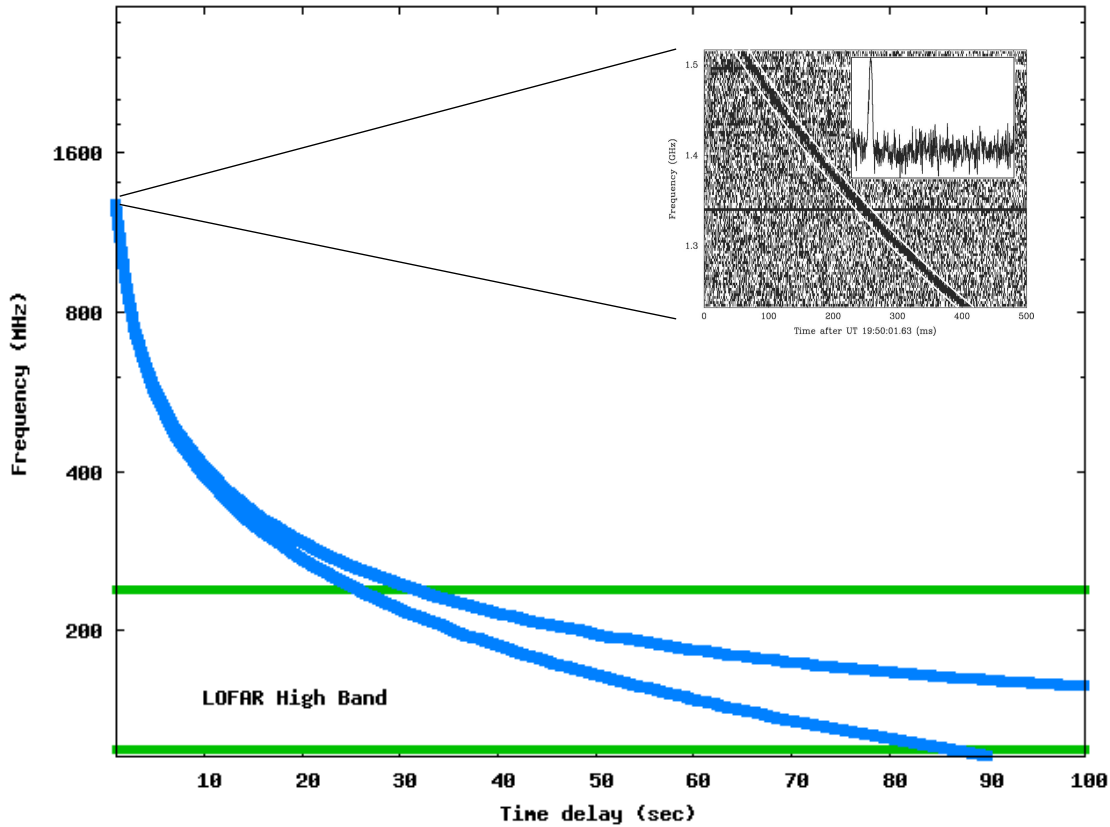}
\caption{An illustration of how the highly-dispersed extragalactic
  radio burst reported by Lorimer et al. (2007) would sweep through
  the LOFAR high band (the frequency limits of which are indicated by
  the green horizontal lines) some tens of seconds later. The blue
  lines indicate the estimated delay and width of the pulse. The
  dispersion delay is assumed to be quadratic; the scatter-broadening
  of the signal is assumed to grow as $\nu^{-4}$ and we assume a pulse
  width of 5 ms at 1.4 GHz (note that the pulse was not resolved by
  Lorimer et al. and so this may be considered to be an {\em upper
  limit} to the scattered pulse width). If the measured steep spectrum
  ($S_{\nu} \propto \nu^{-4}$, the same as the scatter broadening)
  extends to low frequencies, such a burst would be detectable in the
  LOFAR standard data products up to distances in excess of a Gpc,
  allowing unprecedented studies of dispersion and scattering in the
  IGM. Such an event may even have been associated with a neutron
  star--neutron star merger. If so, such events may be detectable by
  LIGO, and a distance inferred from the gravitational wave signal
  alone. LOFAR identification of the host galaxy, via precise
  localisation of the burst, would allow two independent measurements
  of distance on cosmological scales, providing a unique test of
  gravity and of the distance--redshift relation.}
\end{figure}

\item{{\em Flare stars, brown dwarfs and active binaries} are likely to be
  present in almost every LOFAR beam, giving off highly
  circularly-polarised radio bursts from coherent emission processes.
  Potential targets include M dwarf flare stars, active binaries like
  RS CVn and low mass L and T dwarfs.}

\item{LOFAR will also study radio emission from {\em planets} both within
  and beyond the Solar System.  This includes imaging Jupiter's
  magnetosphere at high spatial and time resolution, imaging Jupiter's
  radiation belts, and studying planetary lightning from the other
  planets within the Solar System.  It is also predicted that radio
  bursts from nearby so-called `hot Jupiter' {\em exoplanets} might also be
  detected, and we will carry out a detailed survey for such
  objects. If successful, we would have an inclination independent
  catalogue of extrasolar planets, including possible diagnostics of
  their magnetic fields and rotation periods.}

\item{LOFAR may detect {\em extragalactic radio bursts}, such as that
  reported by Lorimer et al. (2007; Fig 2), to very large distances,
  possibly as far as $z \sim 7$, providing a unique probe of the
  properties of the intergalactic medium (via their dispersion
  measure). Such events may be even be associated with neutron
  star--neutron star mergers, in which case the radio identification
  of events detectable by advanced LIGO may be possible. Since such
  mergers are predicted to provide independent distance measurements
  based on their gravitational wave signatures, identification of the
  electromagnetic counterpart would provide a unique test of gravity
  on cosmological scales, as well as an independent test of the
  redshift-distance relation.}

\end{itemize}

\begin{figure}
\includegraphics[width=.99\textwidth]{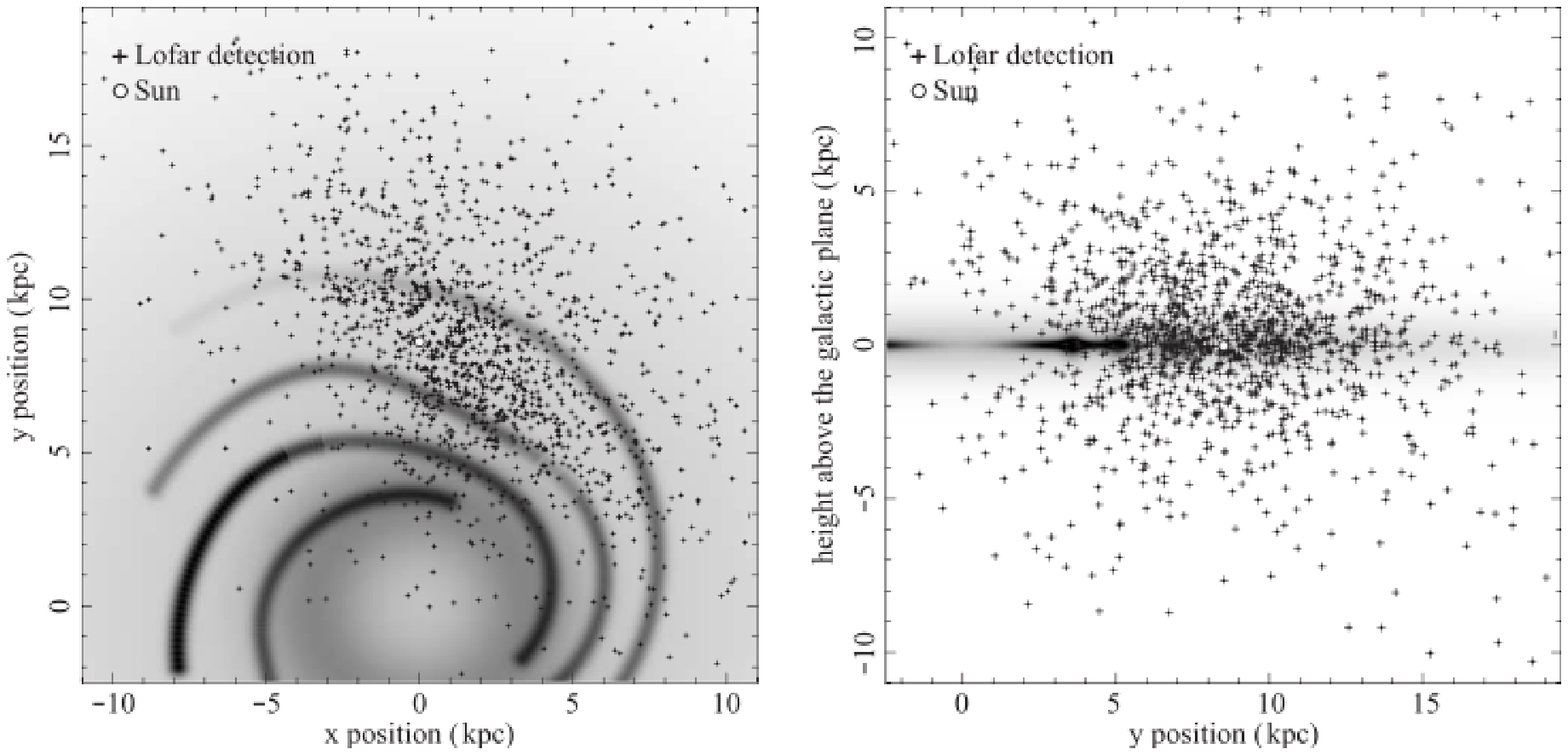}
\caption{ The 1000+ pulsars discovered in a 60-day LOFAR all-sky
        survey simulation.  ISM shown in gray. Left) projected on the
        Galactic plane. Right) projected on the plane through the
        Galactic centre and sun, perpendicular to the disk.  From van
        Leeuwen \& Stappers (2008).}
\end{figure}

In addition to all this, a major survey of classical radio pulsars
will be undertaken, as well as the study of related objects such as
Anomalous X-ray Pulsars (AXPs) and Rotating Radio Transients
(RRATs). The LOFAR pulsar survey is expected to discover more than
1000 new pulsars (see Fig 3), which will provide the majority of
pulsars for the {\em Pulsar Timing Array} (Foster \& Backer 1990) in
the northern hemisphere. Such a survey also has a fair chance of
turning up the first pulsar -- black hole binary. In addition, LOFAR
will provide the sensitivity to allow us to study the individual
pulses from an unprecedented number of pulsars including millisecond
pulsars and the bandwidth and frequency agility to study them over a
wide range which will provide vital new input for models of pulsar
emission. This will provide us with an unparalleled survey of the
population of massive star end-products within our galaxy.

\begin{figure}
\includegraphics[width=.99\textwidth]{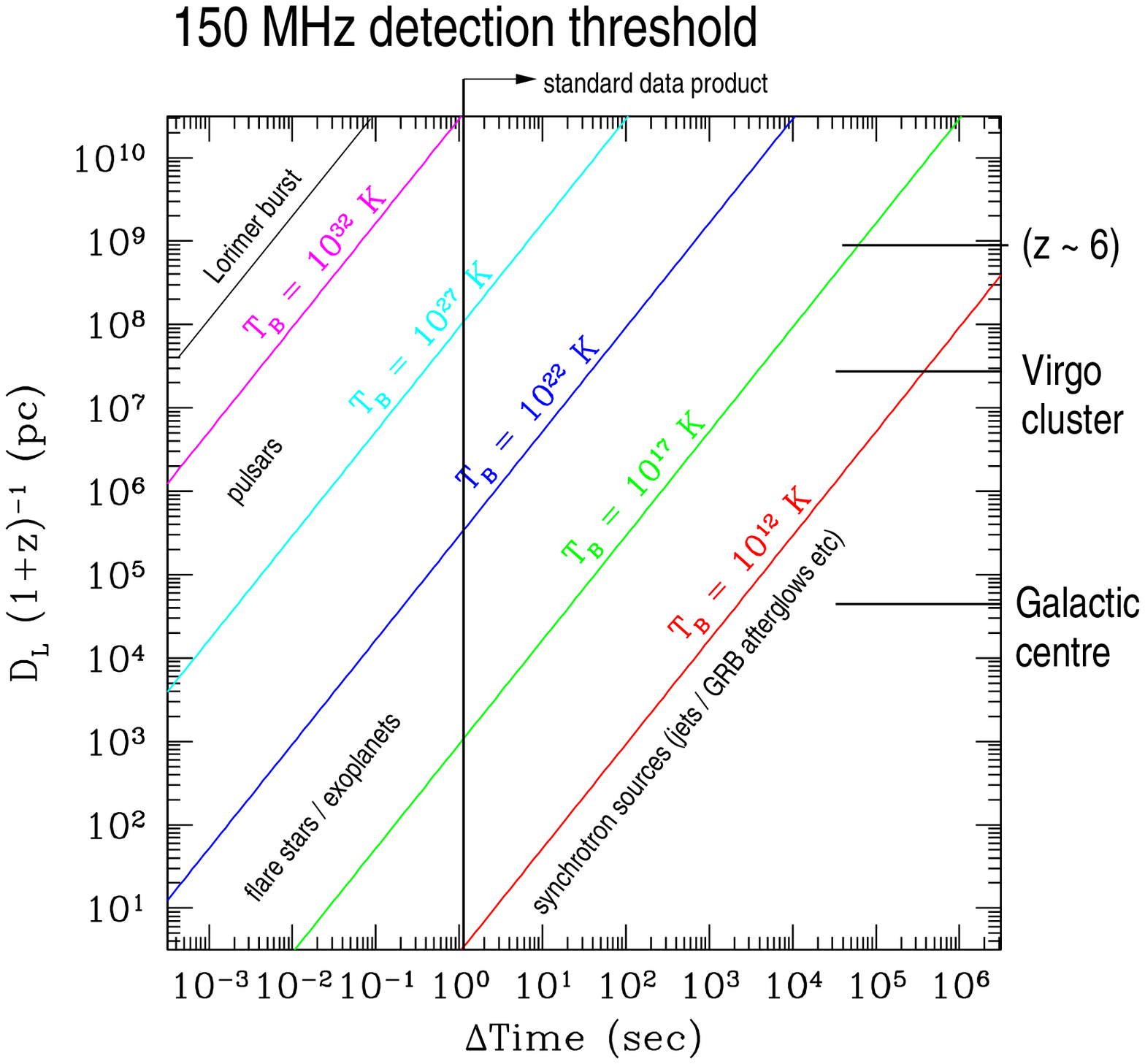}
\caption{An illustration of discovery space possibilities for LOFAR
  when searching for variable and transient phenomena. The diagonal
  lines delineate the upper bound to the region below which detection
  is possible at a given distance and on a given timescale. Also
  indicated are the distances of the galactic centre, Virgo cluster
  and a redshift of $z \sim 6$. Variable Synchrotron sources at all
  distances will be detectable with the standard LOFAR data products,
  which are maps every second. Higher brightness temperature phenomena
  such as flare stars and, we predict, exoplanets are likely to show
  temporal structure on much shorter timescales and so in order to
  obtain the most astrophysical information higher time resolution
  modes will be required. Very high brightness temperature phenomena
  such as pulsars or the `Lorimer transient' (Lorimer et al. 2007)
  could be detected at enormous distances and in many cases will be
  dispersed over timescales long enough to allow their identification
  in standard data products (see e.g. Fig 2). The detection thresholds
  are estimated for a 32 MHz bandwidth.}
\end{figure}

Finally, beyond all this, LOFAR is almost guaranteed to produce many
exciting physical discoveries simply because of the enormous range of
parameter space it is exploring. This could include {\em SETI}, for
which, depending on the nature of the signal, LOFAR may be the best
hope prior to the SKA. This discovery space, and the distance to which
LOFAR may detect different phenomena with different timescales, is
illustrated in Fig 4.

\end{enumerate}

More information on the Transients KSP can be found at

\smallskip
{\bf http://www.astro.uva.nl/lofar\_transients/}
\smallskip

\section{The Radio Sky Monitor}

The Transients KSP will study variable radio sources via four
distinct approaches:

\begin{enumerate}
\item{We will rapidly and regularly scan a large fraction of the
  entire northern sky in the {\em Radio Sky Monitor} mode. This mode
  is the subject of the majority of the rest of this paper.}
\item{We will undertake surveys in a phased-array mode, allowing
  e.g. high time resolution exoplanet / pulsar searches.}
\item{In {\em piggyback mode} we will attempt to seach {\em all} LOFAR
  observations to look for variable and transient sources, by
  comparing with previous images of that region of sky. }
\item{We will also perform targeted deep / high-resolution
  observations of specific key astrophysical sources, often in concert
  with other facilities -- e.g. orbiting X-ray observatories,
  ground-based optical telescopes.}
\end{enumerate}

\subsection{LOFAR configurations}

LOFAR will employ stations of differing sizes depending on their location:

\begin{itemize}
\item{{\bf Core} In the LOFAR core, High-Band Antennae (HBAs) will be
  grouped into sub-stations of 24 tiles, two such sub-stations
  associated with one 48-dipole Low-Band Antenna (LBA) station (96
  LBAs are actually deployed, but only 48 can be selected at any given
  time).}
\item{{\bf Intermediate baselines} On intermediate baselines
  (i.e. those within The Netherlands, outside of the core, LOFAR
  stations will be comprised of one 48-tile set of HBAs and one
  48-dipole LBA station (actually 48 of 96, as in Core)..}
\item{{\bf International baselines} On the longest (international)
  baselines, LOFAR stations will consist of 96 HBAs and 96 LBAs}
\end{itemize}

The inhomogenous design outlined above allows LOFAR to optimise both
the imaging quality and survey speed of the telescope for a fixed budget.
Within The Netherlands, it is anticipated that the Core will
correspond to 16 stations (i.e. 768 HBA tiles and 768 LBA dipoles)
within a diameter of $\sim 2$ km. 

\begin{table}
\begin{tabular}{|c|ccccccccc|}
\hline
Configuration & N & $b_{\rm max}$&  $\nu$ & \multicolumn{2}{c}{HPBW} & $N_{2\pi}$ & $\theta$ & $\sigma_{\rm 1 hr}$ & $\sigma_{2\pi}$ \\
              &   &     (km) & (MHz) & (deg) & (sr) & & (arcsec) & (mJy) & (mJy) \\           
\hline
Core & 768+ & 2 & 30 & 9.1 & 0.079 &160 & 1345 & 20 & 50\\
     &     &&120 & 5.3 & 0.027 & 470 & 330 & 0.6 & 2.2 \\
\hline
LOFAR-NL & 1536+ & 100 & 30  & 9.1 & 0.079 & 160 & 25 & 10 & 25 \\
         &           &&120 & 3.7 & 0.013 & 960 & 7 & 0.3 & 1.9 \\
\hline
E-LOFAR & 2000+ & 1000 & 30 & 11.5 & 0.126 & 100 & 1 & 6.5 & 13 \\
        & &&120 & 2.9 & 0.008 & 1560 & 0.4 & 0.2 & 1.6 \\
\hline 
\end{tabular}
\caption{N is total number of HBA tiles or LBA dipoles, $\nu$ is
  observing frequency, HPBW is the half power beam width ($\equiv$
  FWHM) in both degrees and steradians. $N_{2\pi} = (1-\cos({\rm
    HPBW}/\sqrt{2}))^{-1}$ is the number of beams required to tile out
  one entire hemisphere, assuming an offset between pointing centres
  of $(\rm FWHM)/\sqrt{2}$ as used by Condon et al. (1998) in the NVSS
  (see Fig 5 below for example of this configuration for LOFAR
  beams). The angular resolution is $\theta$ and $\sigma_{\rm 1hr}$ is
  the sensitivity (rms) in one hour. The sensitivity of a uniform
  hemispherical survey, conducted over 24hr, is given by
  $\sigma_{2\pi}$.  A 4 MHz bandwidth is assumed. The sensitivity
  scales with observation length as $\sqrt{t}$, similarly with
  bandwidth as $\sqrt{{\rm bandwidth}}$ and linearly with the number
  of antennae $N$. Note that the field of view varies depending on the
  station layout which dominates in each configuration.}
\end{table}

\subsection{Observations}

\begin{figure}
\includegraphics[width=.99\textwidth]{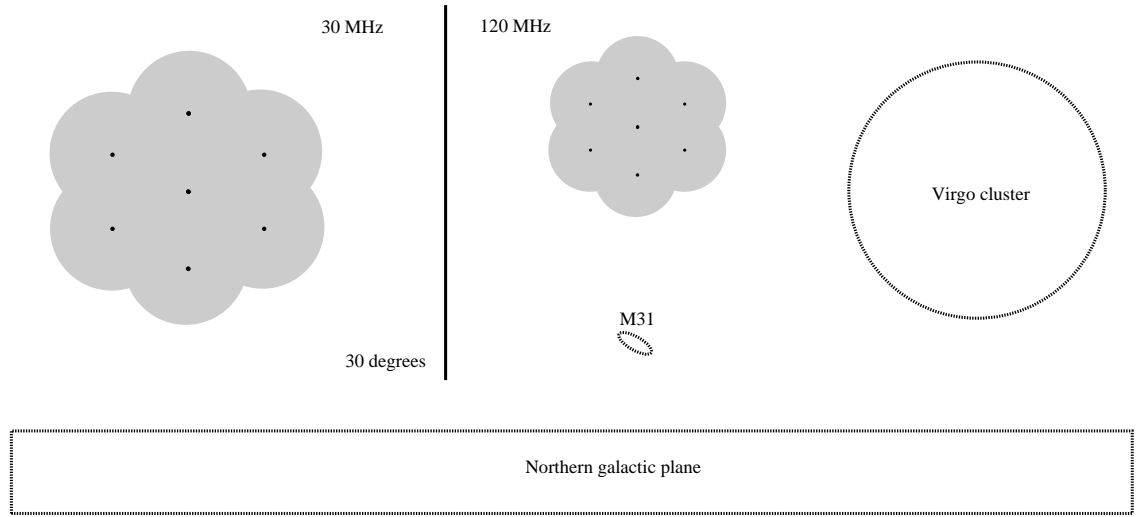}
\caption{An example of a Radio Sky Monitor (RSM) observing modes. In
  this example a hexagonal pattern is traced out using seven beams and
  an offset between pointing centres of ${\rm FWHM}/\sqrt{2} =$6.4 and
  3.7 degrees at 30 and 120 MHz respectively. The approximate angular
  size on the sky of M31, the Virgo cluster and the northern galactic
  plane are indicated for comparison. The angular resolution within
  these fields of view depends upon the longest baselines used -- see
  Table 1.}
\end{figure}

The effective collecting area and field of view of LOFAR peaks at
the lower bound of the two discrete observing bands: 30 MHz for the
LBAs and 120 MHz for the HBA tiles. It is at these two frequencies
that the LOFAR RSM will operate (with possible follow-ups at higher
frequencies).

Specifically, at 30 MHz LOFAR offers the widest possible field of
view.  This is also likely to be the frequency at which coherent
events (which constitute much of the hoped-for exploration of new
parameter space) are strongest. However, spectral index is a key
diagnostic of the nature of a newly-identified source, and a rapid
classification of such events is a key goal of the TKP. In addition,
'synchrotron bubble' events (associated with e.g. GRB afterglows and
some X-ray binary outburst) will peak earlier and stronger at higher
frequencies. In addition, the higher frequency will allow better
localisation of sources, and the HBAs at 120 MHz are more sensitive
than the LBAs at 30 MHz (the factor $\sim 30$ difference in $\sigma$
would be nullified by a spectral index steeper than $\alpha \sim
-2.5$). Finally, dispersion and pulse broadening are strong functions
of frequency, and will be far less severe in the high band than the
low. We therefore propose to operate the RSM at both 30 MHz and 120
MHz.

Using seven beams to tile out a hexagonal pattern (see Fig 5),
operating the LOFAR Radio Sky Monitor from the core (ie. with maximum
baselines of $\sim 2$km) will result in an instantaneous field of view
of $\sim$ 0.08 steradians at 120 MHz and $\sim$ 0.2 steradians at 30
MHz. This hexagonal pattern, with an offset between pointings of
(FWHM)$^{0.5}$ = 6.4 and 3.7 degrees at 30 and 120 MHz respectively,
  gives the most uniform sky coverage for wide field surveys
  (e.g. Condon et al. 1998 for the NRAO VLA sky survey). These fields
  of view are extremely large and illustrate the unprecedented
  monitoring and surveying capabilities of the telescope.

A variety of strategies for operating the RSM mode can be envisaged, e.g.

\begin{itemize}
\item{{\bf Rapid All-Sky Monitoring}: Rapid shallow half-sky
  (hemispherical) surveys could be performed on short timescales in
  order to survey for rapid transients. In table 1, $N_{2\pi}$
  corresponds to the number of beams required to tile out an entire
  hemisphere on the sky. This varies from of order 100 at 30 MHz to
  $\sim 1600$ for 120 MHz with international stations (which are
  larger and therefore have a smaller field of view). This is a very
  small number of pointings in order to survey half of the entire sky,
  and means that several minutes could be spent at each point in the
  sky (if such a scan were carried out over something like one day).
  This in turn means that $\sim$ mJy sensitivity (see $\sigma_{2\pi}$
  in table 1) surveys can be carried out daily.  }
\item{{\bf Zenith monitoring}: Staring at the zenith optimises the
  sensitivity and beam stability of the telescope, whilst providing a
  sizeable and repeatedly monitored part of the sky.  Using the
  patterns illustrated in Fig 5, mapping out the entire field of view
  which passes the zenith can be achieved in approximately 20
  pointings at 30 MHz and 30 at 120 MHz, tracking each field for about
  an hour, achieving (sub-)mJy sensitivity. In RA, Dec terms this
  delineates a strip at 54$^{\circ}$ north of full width 26 degrees
  (i.e. extending from Dec +41 to +67).}
\item{{\bf Galactic plane monitoring}: Most of the northern galactic
  plane is visible from LOFAR and could be monitored for galactic
  transients.  A large fraction of the `known' radio transients, such
  as those associated with rotation-powered neutron stars and X-ray
  binary systems, are heavily concentrated towards the galactic plane
  (note that many sources e.g. flare stars, GRB afterglows, AGN are
  not). The RSM zenith-monitoring mode will sample the galactic plane
  between about $90 \leq l \leq 160$. However, much of the galactic
  plane will not be well sampled by this program, especially towards
  the galactic centre. In order to counterbalance this, we propose
  regular scans of the galactic plane. Since the galactic plane will
  also, however, suffer most from strong field sources and heavy
  dispersion at 30 MHz, these surveys will only be conducted at 120
  MHz. }
\end{itemize}

Fig 6 sketches out the hemispherical and zenith-monitoring modes.

\begin{figure}
\includegraphics[width=.99\textwidth]{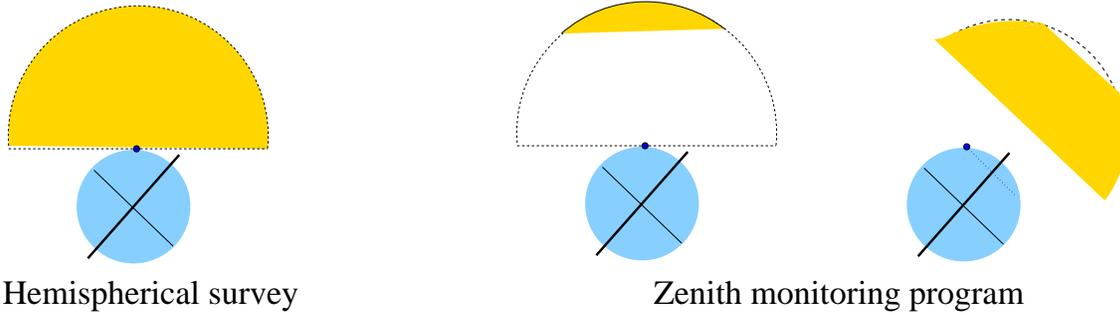}
\caption{Possible modes for observing with the LOFAR Radio Sky Monitor
(RSM). Rapid all-hemisphere surveys are possible (see $N_{2\pi}$ and
$\sigma_{2\pi}$ in Table 1. Alternatively, we may stare at the zenith,
where the array is most sensitive and best calibrated, and allow
rotation of the Earth to tile out a significant fraction of the sky.
}
\end{figure}

\subsection*{Analysis}

Our aim is that the data will be correlated, calibrated and delivered
to Amsterdam at a rate of one image per beam per second, corresponding
to the LOFAR `standard data products'. Delivery of maps made from
longer, logarithmically-spaced time intervals, from the central
computing cluster, will result in a total image transfer rate of
approximately two per beam per second, i.e. 48 images / second for the
RSM. Images will be analysed in real time (via source-detection
software currently under development by the TKP) to check both for new
transients and for the fluxes of known objects. Of course non-standard
modes, such as phase-arrays for pulsar and exoplanet searches, will
require different routes to transient discovery and classification.

Any new transients will have their data fed into a classification /
alert pipeline which may under certain circumstances trigger follow up
observations (such as full-array and frequencies up to 240 MHz). A
sketch of how such a pipeline may integrate with the LOFAR central
processing is presented in Fig 7. {\em It will be our policy that
  alerts will passed to the broader community as well as directly to
  partner observatories}. The fluxes of known variable sources will be
fed into the transients database (currently in the design phase) and
may also trigger follow-up observations if there are particularly
dramatic changes.

\begin{figure}
\includegraphics[width=.95\textwidth]{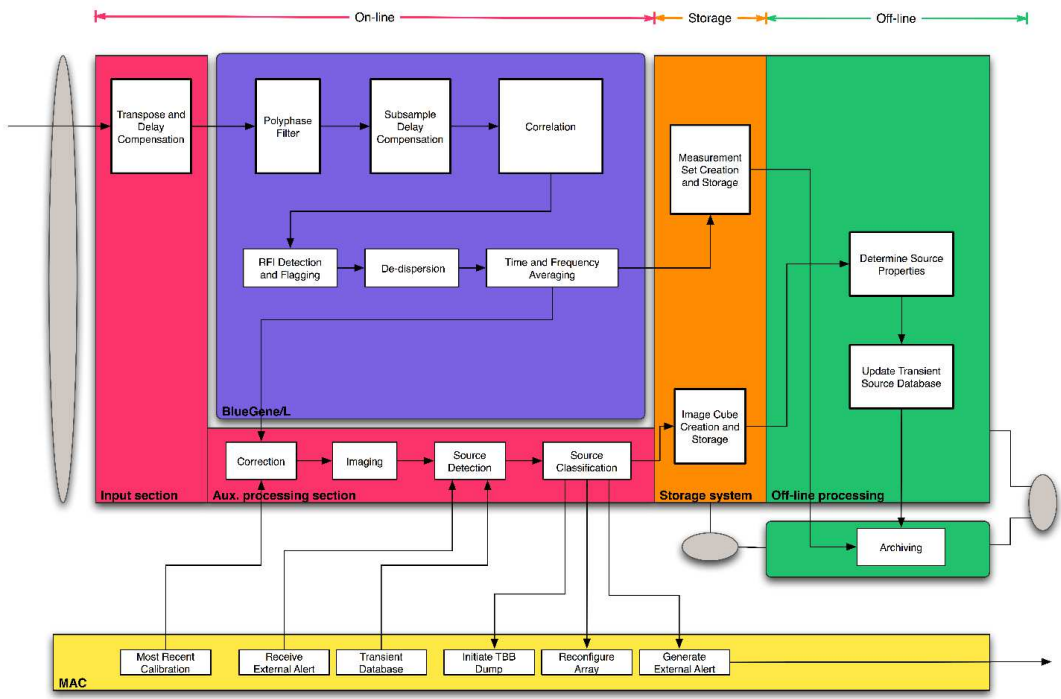}
\caption{An simple illustration of how transients pipelines are to be
  integrated into the processing system for LOFAR.}
\end{figure}


\begin{thebibliography}{99}

\bibitem{}
Condon J.J., Cotton W.D., Greisen E.W., Yin Q.F., Perley R.A., Taylor
G.B., Broderick J.J., 1998, AJ, 115, 1693
\bibitem{}
Foster R.S., Backer D.C., 1990, ApJ, 361, 300 
\bibitem{}
Gunst A., van der Schaaf K., Bentum M.J., published in the Proceedings
from SPS-DARTS 2006, The second annual IEEE BENELUX/DSP Valley Signal
Processing Symposium, March 28-29 (2006), Antwerp, Belgium
\bibitem{}
Lorimer D.R., Bailes M., McLaughlin M.A., Narkevic D.J., Crawford F.,
2007, Science, 318, 777
\bibitem{} 
van Leeuwen J., \& Stappers B., 2008, published in 40 YEARS OF
PULSARS: Millisecond Pulsars, Magnetars and More. AIP Conference
Proceedings, Volume 983, pp. 598-600 (2008) {\bf (arXiv:0710.0675)}



\end{thebibliography}
\end{document}